\documentclass{article}

\usepackage{arxiv}

\usepackage[utf8]{inputenc}
\usepackage[T1]{fontenc}
\usepackage{hyperref}
\usepackage{url}
\usepackage{booktabs}
\usepackage{amsfonts}
\usepackage{nicefrac}
\usepackage{microtype}
\usepackage{graphicx}
\usepackage{natbib}
\usepackage{doi}
\usepackage[USenglish]{babel}
\usepackage[intlimits]{amsmath}
\usepackage{bm}
\hypersetup{colorlinks=true, citecolor=blue}
\usepackage{amssymb}
\usepackage{fontawesome}
\usepackage{mathtools}
\usepackage{xcolor}
\usepackage{lmodern}
\usepackage{siunitx}
\usepackage{booktabs}
\usepackage{etoolbox}

\renewrobustcmd{\bfseries}{\fontseries{b}\selectfont}
\renewrobustcmd{\boldmath}{}
\newrobustcmd{\B}{\bfseries}

\title{The Case against Generally Weighted Moving Average (GWMA) Control Charts}

\date{June 30, 2021}

\author{ \href{https://orcid.org/0000-0002-9666-5554}{\includegraphics[scale=0.06]{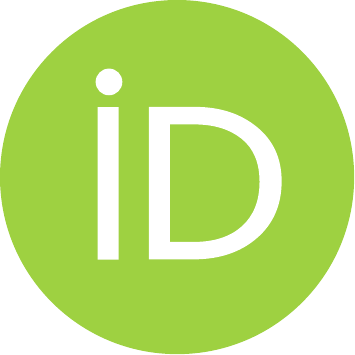}\hspace{1mm}Sven Knoth} \\
	Dep. of Mathematics \& Statistics\\
	Helmut Schmidt University\\
	Hamburg, Germany\\
	\texttt{knoth@hsu-hh.de} \\
	\And 
	\href{https://www.stat.vt.edu/people/stat-faculty/woodall-bill.html}{\faHome\hspace{1mm}William H. Woodall} \\
	Dep. of Statistics\\
	Virginia Tech\\
	Blacksburg VA, USA\\
	\texttt{bwoodall@vt.edu} \\
	\And
	\href{https://research.tec.mx/vivo-tec/display/PID_77538}{\faHome\hspace{1mm}V{\'i}ctor G. Tercero-G{\'o}mez} \\
	School of Engineering \& Sciences\\
	Tecnologico de Monterrey\\
	Monterrey, Nuevo Leon, Mexico\\
	\texttt{victor.tercero@tec.mx} \\
}

\renewcommand{\shorttitle}{The Case against GWMA Charts}

\begin{document}
	
\maketitle

\begin{abstract}
We argue against the use of generally weighted moving average (GWMA) control charts. Our primary reasons are the following: 1) There is no recursive formula for the GWMA control chart statistic, so all previous data must be stored and used in the calculation of each chart statistic. 2) The Markovian property does not apply to the GWMA statistics, so computer simulation must be used to determine control limits and the statistical performance. 3) An appropriately designed, and much simpler, exponentially weighted moving average (EWMA) chart provides as good or better statistical performance. 4) In some cases the GWMA chart gives more weight to past data values than to current values.
\end{abstract}


\section{Introduction}

The generally-weighted moving average (GWMA) control chart was introduced by
\cite{Sheu:Lin:2003} with a follow-up paper by \cite{Sheu:Yang:2006a}.
\cite{Mabu:EtAl:2021} provided a full, up-to-date review of the literature on the GWMA methodology, which currently includes at least sixty-six papers. 

As the name indicates, the GWMA chart is based on statistics which are weighted averages of the observations collected over time. It is a generalization of the exponentially weighted moving average (EWMA) chart for which the weights decrease exponentially with the age of the observations.

We see serious disadvantages of the GWMA approach, with no redeeming benefits. In particular there is no recursive formula for the control chart statistic, so all previous data must be stored and used in each control chart statistic calculation. The Markovian property does not apply, so computer simulation must be used to determine appropriate control limits as well as the statistical performance. The EWMA chart does not possess these disadvantages. It has been argued that the GWMA control chart is justified because it has better statistical performance than the EWMA chart. We show, however, that the comparisons have been flawed. An appropriately designed, and much simpler, EWMA chart, can provide as good or better statistical performance for any given GWMA chart.

\section{GWMA and EWMA Charts}

Let $X_1, X_2, \ldots$ be a sequence of independent and identically distributed (i.i.d.) random variables that follow a $\mathcal{N}(\mu,\sigma^2)$ distribution. At time $t$ the GWMA charting statistic is defined as
\begin{align}
  G_t & = P(M=1) X_t + P(M=2) X_{t-1} + \ldots + P(M=t) X_1 + P(M>t) \mu_0  \nonumber \\
  & = \left(q^{0^\alpha} - q^{1^\alpha}\right) X_t
    + \left(q^{1^\alpha} - q^{2^\alpha}\right) X_{t-1} + \ldots
    + \left(q^{(t-1)^\alpha} - q^{t^\alpha}\right) X_1 + q^{t^\alpha} \mu_0
  \nonumber \\    
  & = \sum_{i=1}^t \left(q^{(i-1)^\alpha} - q^{i^\alpha}\right) X_{t-i+1}
  + q^{t^\alpha} \mu_0 \,, \label{eq:01}
\end{align}
where $P(M=i)$ is said to be the probability of an out-of-control signal at time $i$. Here $\mu_0$ represents the in-control mean, and the design parameters $q$ and $\alpha$ ($0<q<1\,,\;\alpha>0$) determine the weighting structure. If dealing with subgroups of size $n > 1$, then one would use the sequence of sample means $\bar{X}_i$ each with variance $\sigma^2/n$. For simplicity, we assume that the mean of the process can change, but that the standard deviation is known and remains constant at $\sigma_0$.

If the process is in-control, the expected value and variance of $G_t$ are
\begin{align*}
    E(G_t) & = \mu_0 \,, \\
    Var(G_t) & = Q_t \sigma_0^2 \,,
\end{align*}
where
\begin{equation}
  Q_t = \sum_{i=1}^t \left(q^{(i-1)^\alpha} - q^{i^\alpha}\right)^2 \,. \label{eq:02}
\end{equation}
A process is said to be out-of-control once $G_t$ falls outside the control limits
\begin{equation*}
  \mu_0 \pm L_G \sqrt{Q_t} \sigma_0 \,.     
\end{equation*}
As $t$ gets larger, $Q_t$ approaches a limit, with no closed form, of
\begin{equation}
  Q = \lim\limits_{t\to\infty} \sum_{i=1}^t \left(q^{(i-1)^\alpha} - q^{i^\alpha}\right)^2, \label{eq:03}      
\end{equation}
and the control limits simplify to
\begin{equation*}
  \mu_0 \pm L_G \sqrt{Q} \sigma_0 \,.     
\end{equation*}
The control limit constant $L_G$ is determined in order to meet a specified in-control performance. The weights are defined as probabilities, but this is not justified in any way whatsoever. The weights simply form a sequence of non-negative values, typically, but not always, decreasing, that sum to one. This makes the GWMA statistic $G_t$ a convex combination of the in-control mean $\mu_0$ and all monitored observations. If $\alpha = 1$, then the GWMA chart reduces to an EWMA chart with $\lambda = 1-q$. In Figure 1 we present the weights, expanding
\begin{figure}[hbt]
\centering
\includegraphics[width=.7\textwidth]{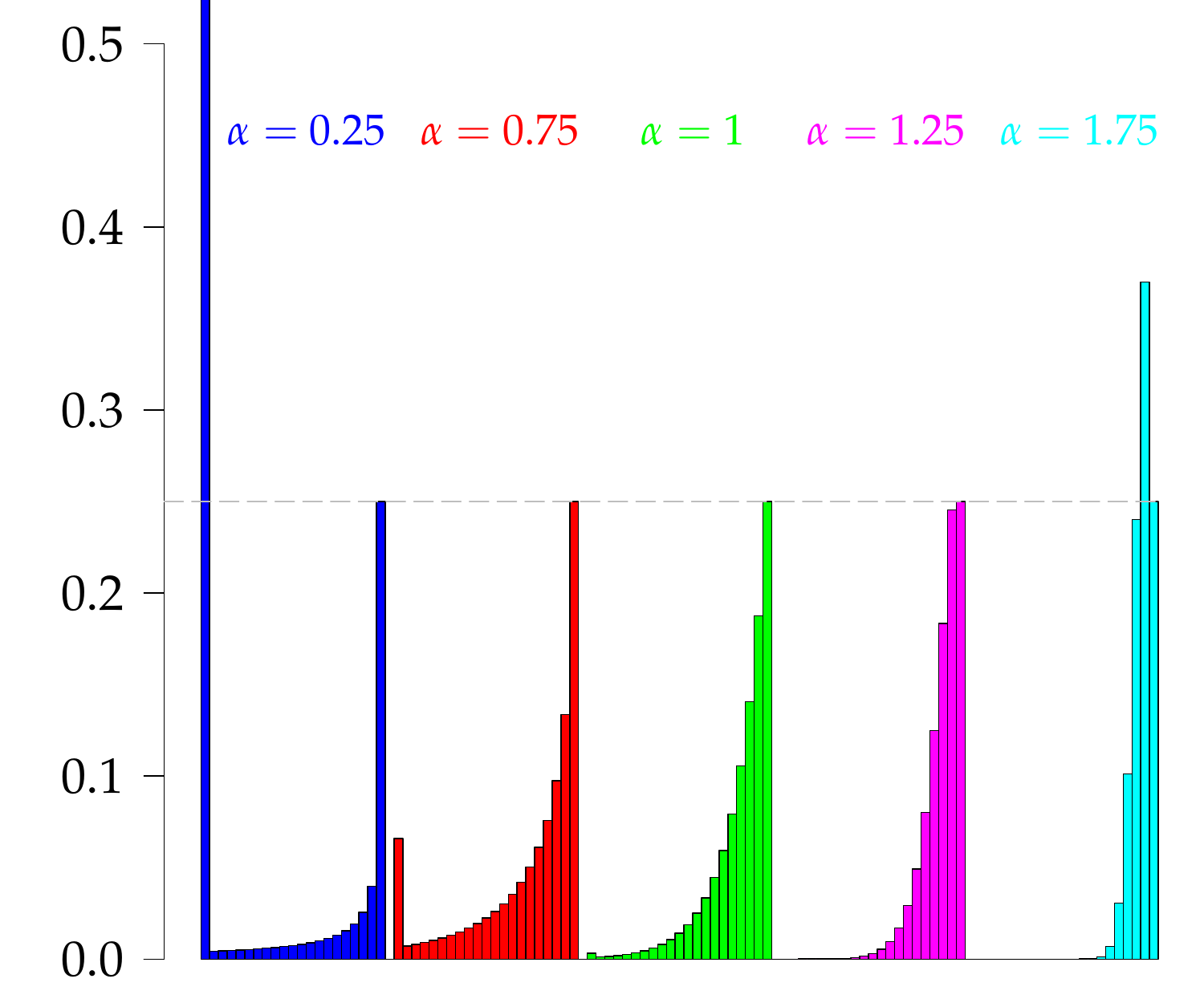}
\caption{Extended versions of Figure A3 for case $q = 0.75$.}\label{fig:weightsA3}
\end{figure}
Figure A3 in \cite{Mabu:EtAl:2021}, where for five selected values of $\alpha$ all weights including at $\mu_0$ are presented for $t = 20$. The weight for the most recent observation is on the right. We note from the plot on the right-hand-side of Figure 1 that the weights do not always decrease with the age of the observations (e.\,g., for $\alpha = 1.75$).

\cite{Luca:Sacc:1990a} and others have studied the EWMA chart in detail. The EWMA chart statistics, along with the in-control mean and variance, are
\begin{align*}
  Z_t & = (1-\lambda) Z_{t-1} + \lambda X_t \nonumber \\
  & = \lambda \sum_{i=1}^t (1-\lambda)^{i-1} X_{t-i+1}
  + (1-\lambda)^t \mu_0 \,, \nonumber \\
  E(Z_t) & = \mu_0 \,, \nonumber \\
  Var(Z_t) & = \sigma^2_0 \frac{ \lambda}{2-\lambda} \big( 1 - (1-\lambda)^{2t} \big).
\end{align*}

It is easy to see that, as $t\to\infty$, the scaled variance  $Var(Z_t)$ approaches a constant, i.e,
\begin{equation} \label{eq:04}
    Var(Z_t)/\sigma_0^2 \;\to\;  Q_E = \frac{\lambda}{2-\lambda} \,.
\end{equation}

We note that there is no recursive formula for the computation of $G_t$, $t = 1, 2, 3, \ldots$ so all previous data must be stored and used in the calculations required by Eq. \eqref{eq:01}. Clearly the EWMA chart has the advantage in this respect. In addition, the successive values of the EWMA statistic satisfy the Markov property, leading to accurate analytical methods based on Markov chains and integral equations for determining the statistical properties of the chart, as discussed by \cite{Luca:Sacc:1990a} and others. Here we will utilize
results given in \cite{Knot:2003, Knot:2005a}, who extended the integral equation method to varying control limits, in particular, for
$\mu_0 \pm L_E \sqrt{Var(Z_t)}$.
A readily available implementation is the function \texttt{xewma.arl()} provided by the \textsf{R} package \texttt{spc} \citep{Knot:2021}.

\section{ARL Comparisons}

The added computations and complexity of the GWMA chart have been typically justified based on average run length (ARL) comparisons. The following Table 1, for example, was taken from \cite{Sheu:Lin:2003}.
\begin{table}[hbt]
\centering
\caption{ARL values for some GWMA charts with $q=0.75$ from Table 1 in \cite{Sheu:Lin:2003}.}\label{tab:01}
\begin{tabular}{crrrrr} \toprule
  & \multicolumn{5}{c}{$q=0.75$} \\ \midrule
  & $\alpha = 0.50$ & $\alpha = 0.75$ & $\alpha = 0.80$ & $\alpha = 0.90$ & $\alpha = 1.00$ \\
  $\frac{\mu-\mu_0}{\sigma_0}$ &
  $L_G=3.063$ & $L_G=3.028$ & $L_G=3.021$ & $L_G=3.001$ & $L_G=3.002$ \\ \midrule 
 0.00 & 500.63 & 500.17 & 500.36 & 499.99 & 500.00 \\
 0.25 & \B 128.90 & \B 142.60 & \B 147.45 & \B 157.99 & 169.85 \\
 0.50 & \B  40.76 & \B  39.90 & \B  41.09 & \B  43.86 &  47.50 \\
 0.75 &  19.72 & \B  17.76 & \B  17.88 & \B  18.42 &  19.33 \\
 1.00 &  11.74 & \B  10.23 & \B  10.19 & \B  10.20 &  10.43 \\[1ex]
 1.25 &   7.83 &   6.82 &   6.74 & \B   6.67 &  6.68 \\
 1.50 &   5.65 &   4.96 &   4.89 &   4.80 &  4.78 \\
 2.00 &   3.39 &   3.06 &   3.02 &   2.97 &  2.94 \\
 3.00 &   1.73 &   1.65 &   1.64 &   1.63 &  1.62 \\ \bottomrule
\end{tabular}
\end{table}

In Table~\ref{tab:01}, the ARL values corresponding to $\alpha = 1$ are for the competing EWMA chart with $\lambda = 1-q$ = 0.25. As the bolded values indicate, one can achieve quicker detection of small shifts in the process by decreasing the value of $\alpha$. We note, however, that using values $\alpha < 1$ puts more weight on past data, thus leading to the quicker detection of smaller shifts in the mean. A fairer comparison with the EWMA chart would be to reduce the value of the smoothing constant $\lambda$ for the competing EWMA chart. We present such comparisons in Table~\ref{tab:02}, where the GWMA chart no longer shows any advantages in performance.
\begin{table}[hbt]
\centering
\caption{ARL values for some GWMA and competing EWMA charts with $q=0.75$ from Table 1 in \cite{Sheu:Lin:2003} with some further EWMA results.}\label{tab:02}
\begin{tabular}{crrrr|rr} \toprule
  & \multicolumn{4}{c|}{$q=0.75$}
  & \multicolumn{2}{c}{new EWMA} \\ \midrule
  & $\alpha = 0.50$ & $\alpha = 0.75$ & $\alpha = 0.80$ & $\alpha = 0.90$
  & $\lambda = 0.206$ & $\lambda = 0.152$ \\
  $\frac{\mu-\mu_0}{\sigma_0}$
  & $L_G = 3.063$ & $L_G = 3.028$ & $L_G = 3.021$
  & $L_G = 3.001$ & $L_E = 2.971$ & $L_E = 2.915$ \\ \midrule 
 0.00 & 500.63 & 500.17 & 500.36 & 499.99 & 500.00 & 500.00 \\
 0.25 & \B 128.90 & \B 142.60 & \B 147.45 & \B 157.99 & 151.45 & 127.89 \\
 0.50 & \B  40.76 & \B  39.90 & \B  41.09 & \B  43.86 &  41.35 & 34.60 \\
 0.75 & 19.72 & \B  17.76 & \B  17.88 & \B  18.42 &  17.36 & 15.31 \\
 1.00 & 11.74 & \B  10.23 & \B  10.19 & \B  10.20 &   9.68 & 8.90 \\[1ex]
 1.25 &  7.83 &   6.82 &   6.74 & \B   6.67 &  6.36 & 6.00 \\
 1.50 &  5.65 &   4.96 &   4.89 &   4.80 &  4.61 & 4.41 \\
 2.00 &  3.39 &   3.06 &   3.02 &   2.97 &  2.87 & 2.78 \\
 3.00 &  1.73 &   1.65 &   1.64 &   1.63 &  1.60 & 1.56 \\ \bottomrule
\end{tabular}
\end{table}
The two new EWMA configurations in the last two columns of Table~\ref{tab:02} were chosen to achieve the same asymptotic variance, i.\,e., by setting equal $Q = Q_E$ in Eqs. \eqref{eq:03} and \eqref{eq:04},
respectively, for $\alpha = 0.8$ and $\alpha = 0.5$.
It turns out that the zero-state out-of-control ARL values of the
appropriately matched EWMA designs are typically smaller than those of their GWMA counterparts. Even more, the EWMA chart with $\lambda = 0.152$ (the counterpart of $(q=0.75, \alpha =0.5)$) dominates uniformly all considered GWMA designs.

\begin{figure}[hbt]
\centering
\includegraphics[width=.6\textwidth]{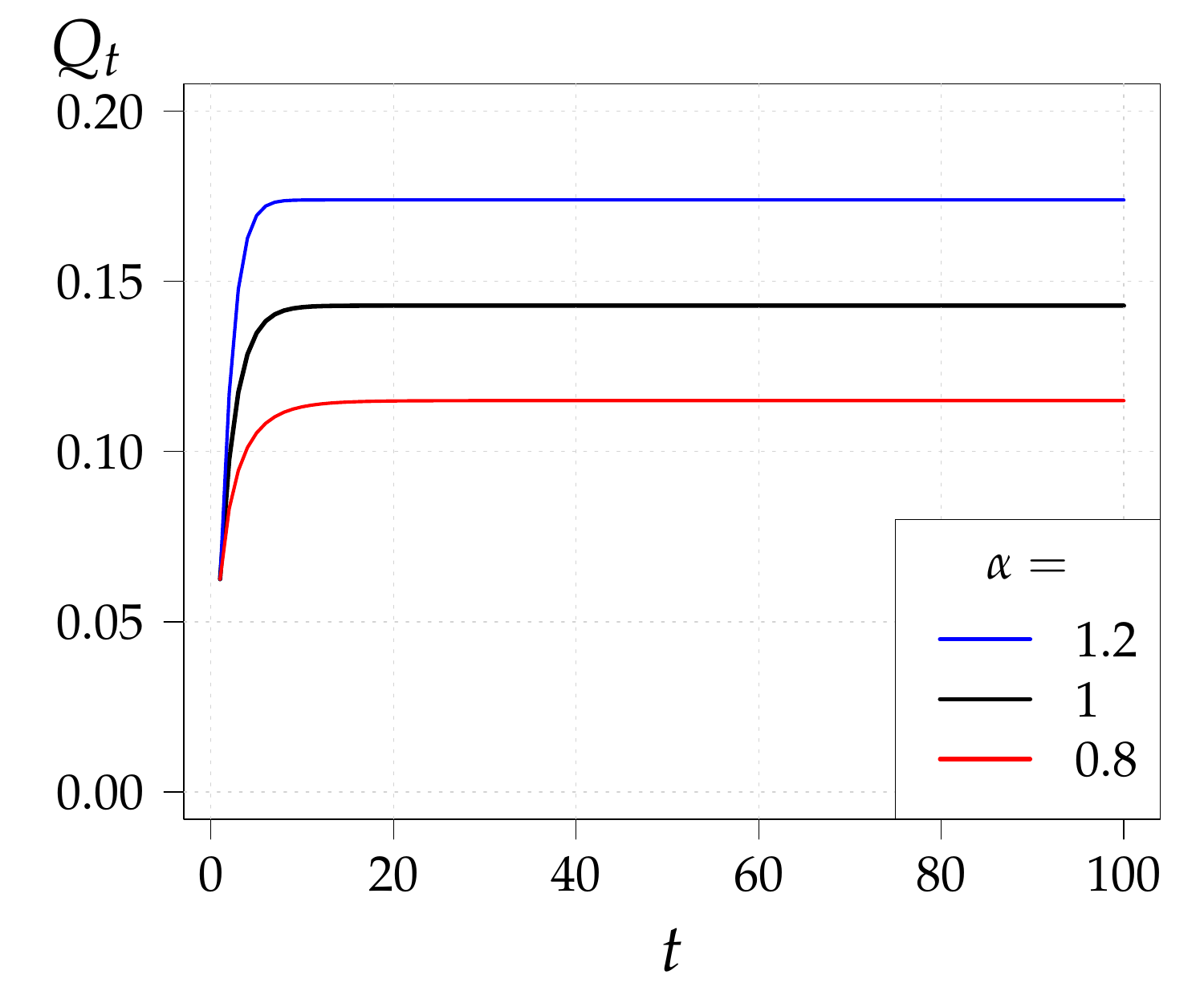}
\caption{GWMA variance $Q_t$ for the first 100 observations with $q = 0.75$, and $\alpha = 0.8$, $=1$ (EWMA), $=1.2$.}\label{fig:Qtseries}
\end{figure}

\begin{figure}[hbt]
\centering
\renewcommand{\tabcolsep}{0mm}
\begin{tabular}{cc}
  \scriptsize Asymptotic GWMA variance $Q$ & \scriptsize Matching $\lambda$ \\[-1ex]
  \includegraphics[width=.48\textwidth]{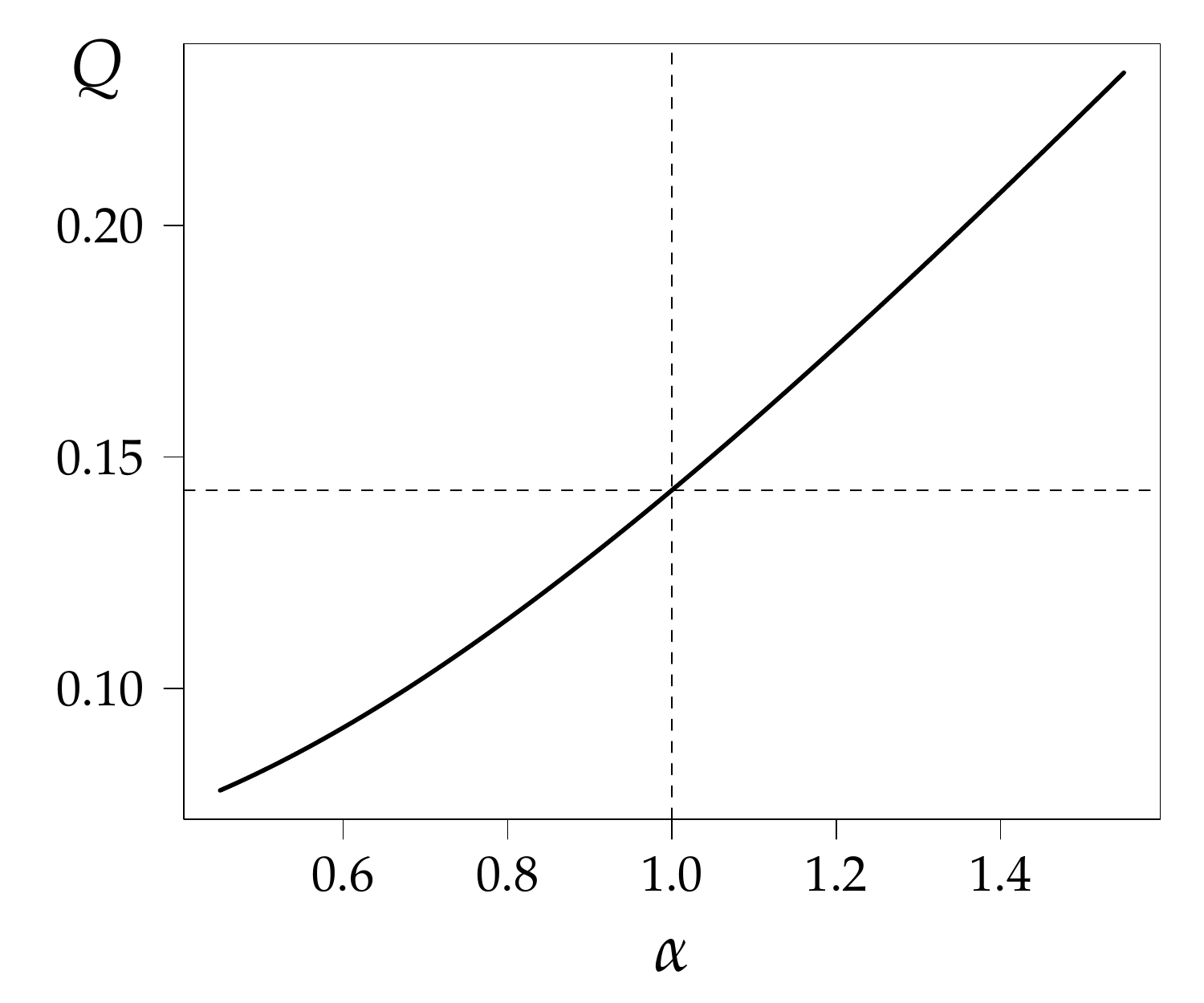} & 
  \includegraphics[width=.48\textwidth]{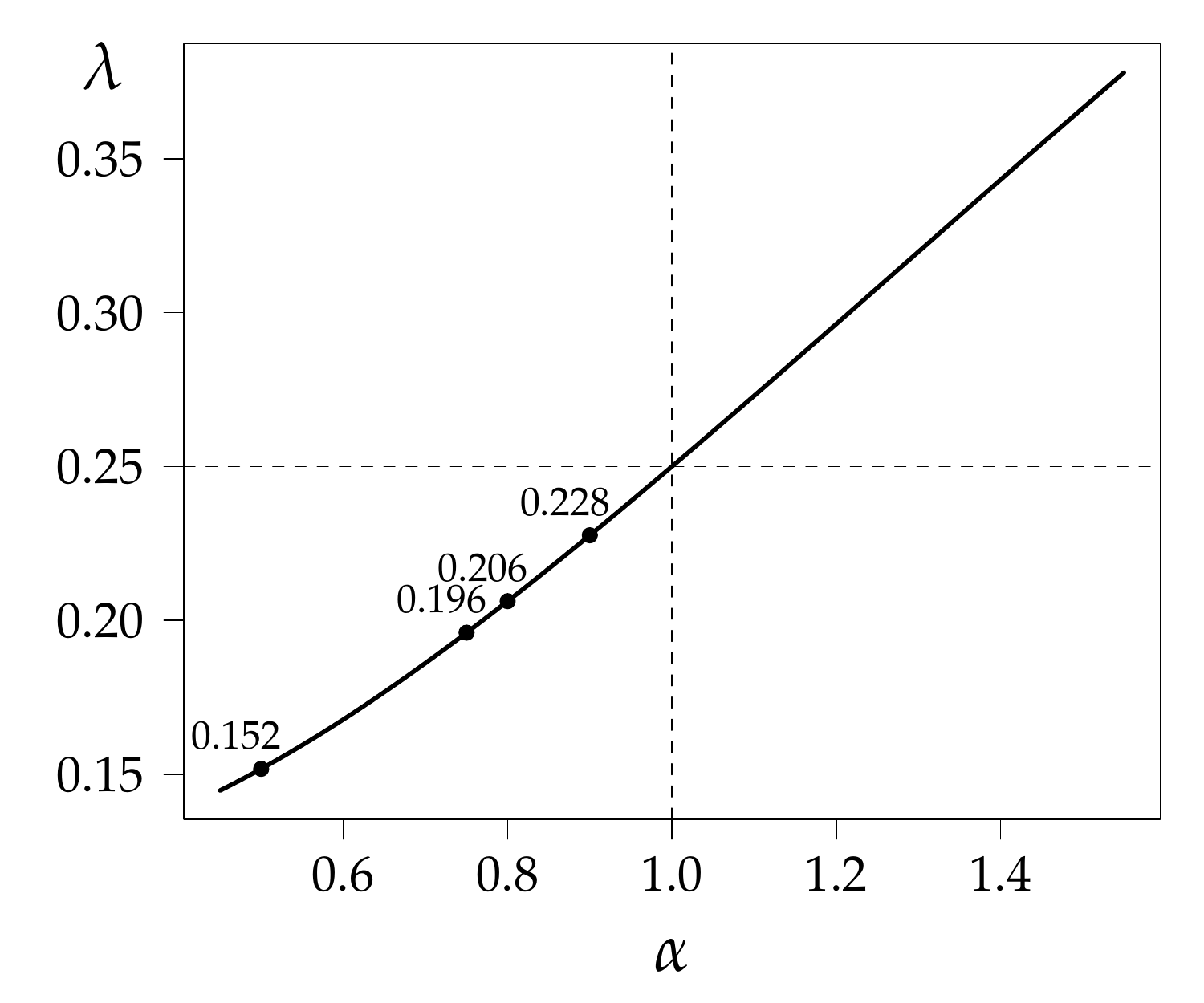}
\end{tabular}
\caption{Asymptotic GWMA variance $Q$ 
and matching $\lambda$ for $q = 0.75$ and $\alpha \in [0.5,1.5]$.}\label{fig:Qlambda_vs_alpha}
\end{figure}

One possibility to match GWMA $(q,\alpha)$ with EWMA $\lambda$ is to equalize their asymptotic variances, as done for the values in Table~\ref{tab:02}. To illustrate the impact of $\alpha$ on $Q$ from Eq.  \eqref{eq:03}, we plot the first 100 values of  $Q_t$ for $q = 0.75$ and $\alpha \in \{0.8, 1, 1.2\}$ in Figure~\ref{fig:Qtseries}. %
Fortunately, all three series $\{Q_t\}$ converge, but only for the EWMA chart do we have an explicit expression for the limit, cf. to Eq. \eqref{eq:04}. In case of
$\alpha \ne 1$, we utilize $Q_{200}$ as an accurate proxy for the asymptotic value. In Figure~\ref{fig:Qlambda_vs_alpha},
we demonstrate the relationship between $Q$ and $\alpha$,
and provide the EWMA $\lambda$ values so that $Q = Q_E$.

From Table~\ref{tab:02} we conclude that it is not difficult to build EWMA designs which convincingly compete with the corresponding GWMA chart. Following the simple principle to match the asymptotic variances, as illustrated in Figure~\ref{fig:Qlambda_vs_alpha}, yields a reasonable $\lambda$ value. Thus far we have considered the zero-state ARL comparison, i.e., under the assumption that any shift in the mean occurs at the start of monitoring. To get the broader picture, we consider next the conditional expected delay (CED)
and its limit, the conditional steady-state ARL where the shift occurs at the change-point $\tau$. We let $L$ represent the time of the out-of-control signal. The CED and the steady-state ARL are the following:
\begin{align*}
	D_\tau & = E_\tau\big(L-\tau+1\mid L\ge \tau \big) \,, \\
	\mathcal{D} & = \lim_{\tau\to\infty} D_\tau \,.
\end{align*}
As for the zero-state ARL $\mathcal{L}$ we utilize the \textsf{R} package \texttt{spc} to calculate the CED $D_\tau$ and the
steady-state ARL $\mathcal{D}$ of the EWMA chart. These calculations
are much more difficult for the GWMA chart with $\alpha\ne 1$.
\cite{Sheu:Lin:2003} reported that
\textit{``finding the exact ARLs for given control limits is not
straightforward''} because \textit{``the limits of GWMA vary with time''}. The problem is identified, but the cause is different. The series $\{G_t\}$ \textit{``cannot be viewed as following a first-order Markov chain''}, citing \cite{Chak:Huma:Bala:2017}, p. 7796. This basically prevents any accurate Markov chain approximation. In addition, it hinders the Monte Carlo simulation framework substantially. In our Monte Carlo study, we stored up to 10\,000 previous observations to determine the simulation replicates of $G_t$. Moreover, for each setup we generated $10^8$ replications, to estimate $D_\tau$ and $\mathcal{D}$ (here we use $D_{100}$ as substitute). Therefore, all simulated $D_\tau$ profiles appear agreeably smooth.

\begin{figure}[hbt]
\centering
\renewcommand{\tabcolsep}{0mm}
\begin{tabular}{cc}
  \scriptsize $(\mu-\mu_0)/\sigma_0=0.5$ &
  \scriptsize $(\mu-\mu_0)/\sigma_0=1$ \\[-1ex]
  \includegraphics[width=.48\textwidth]{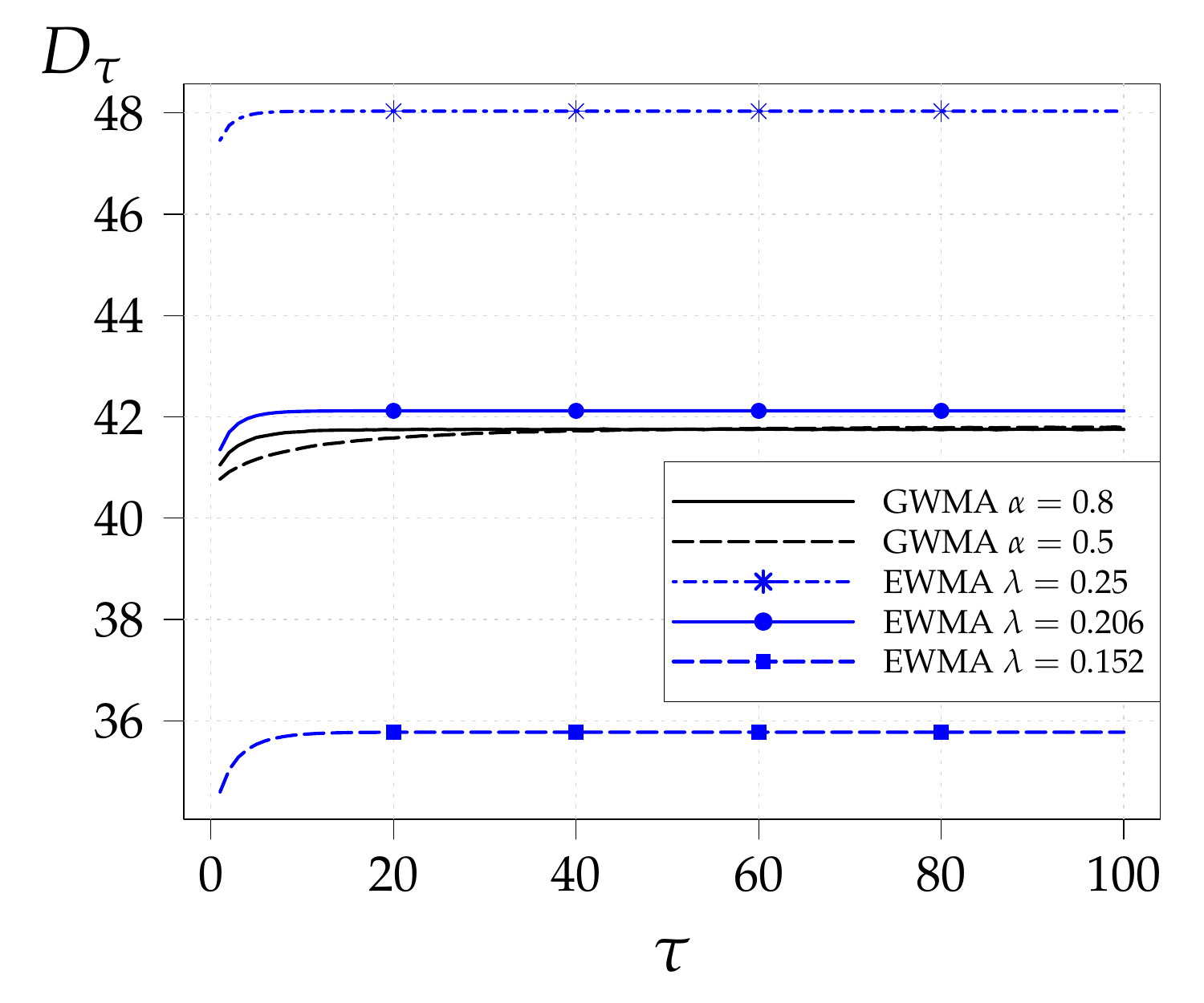} & 
  \includegraphics[width=.48\textwidth]{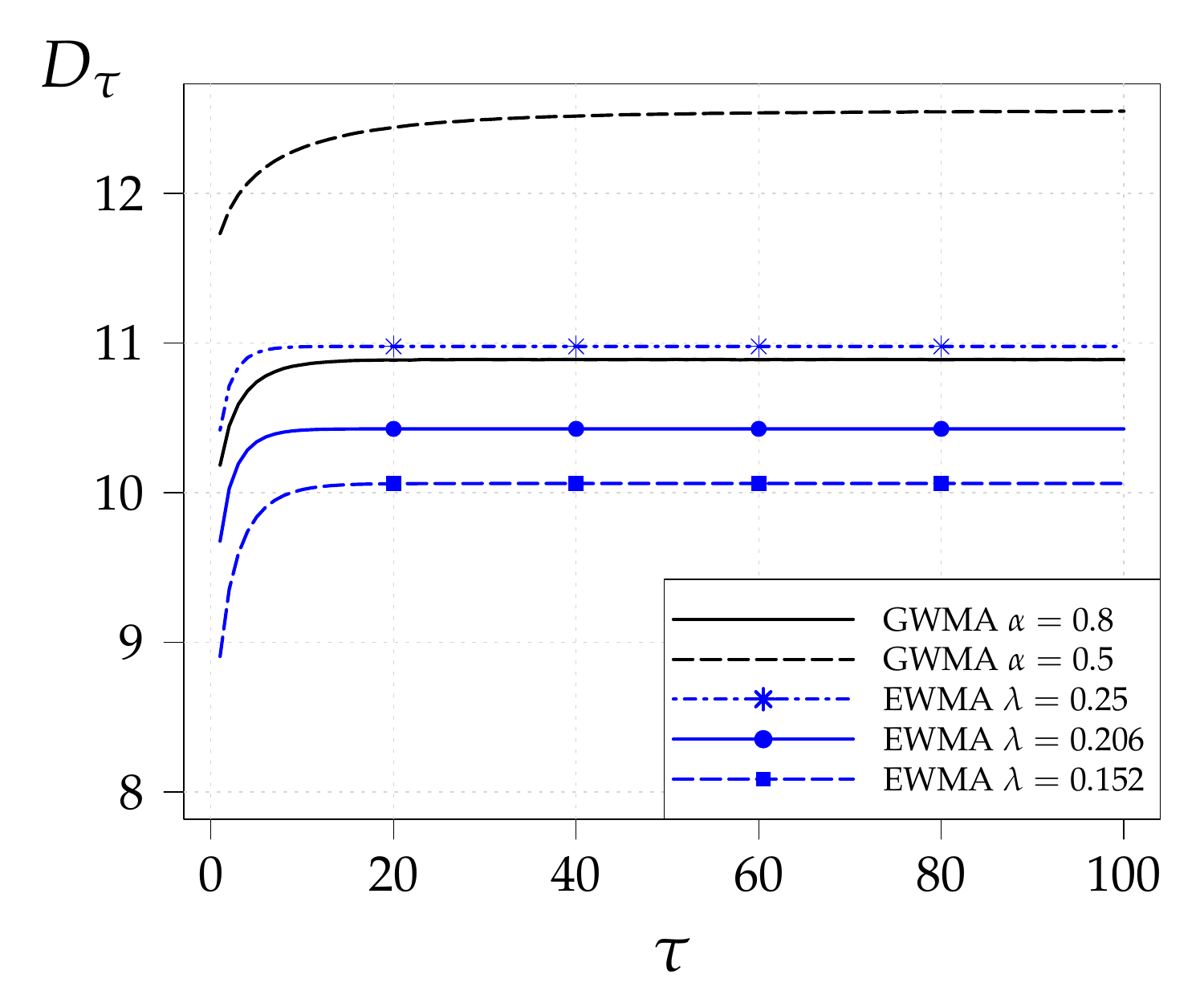}
\end{tabular}
\caption{Conditional expected delay $D_\tau$ for $q = 0.75$, $\alpha \in \{0.5,0.8\}$ and three EWMA designs, small and medium size changes.}\label{fig:dtau_small}
\end{figure}

\begin{figure}[hbt]
\centering
\renewcommand{\tabcolsep}{0mm}
\begin{tabular}{cc}
  \scriptsize $(\mu-\mu_0)/\sigma_0=2$ &
  \scriptsize $(\mu-\mu_0)/\sigma_0 = 3$ \\[-1ex]
  \includegraphics[width=.49\textwidth]{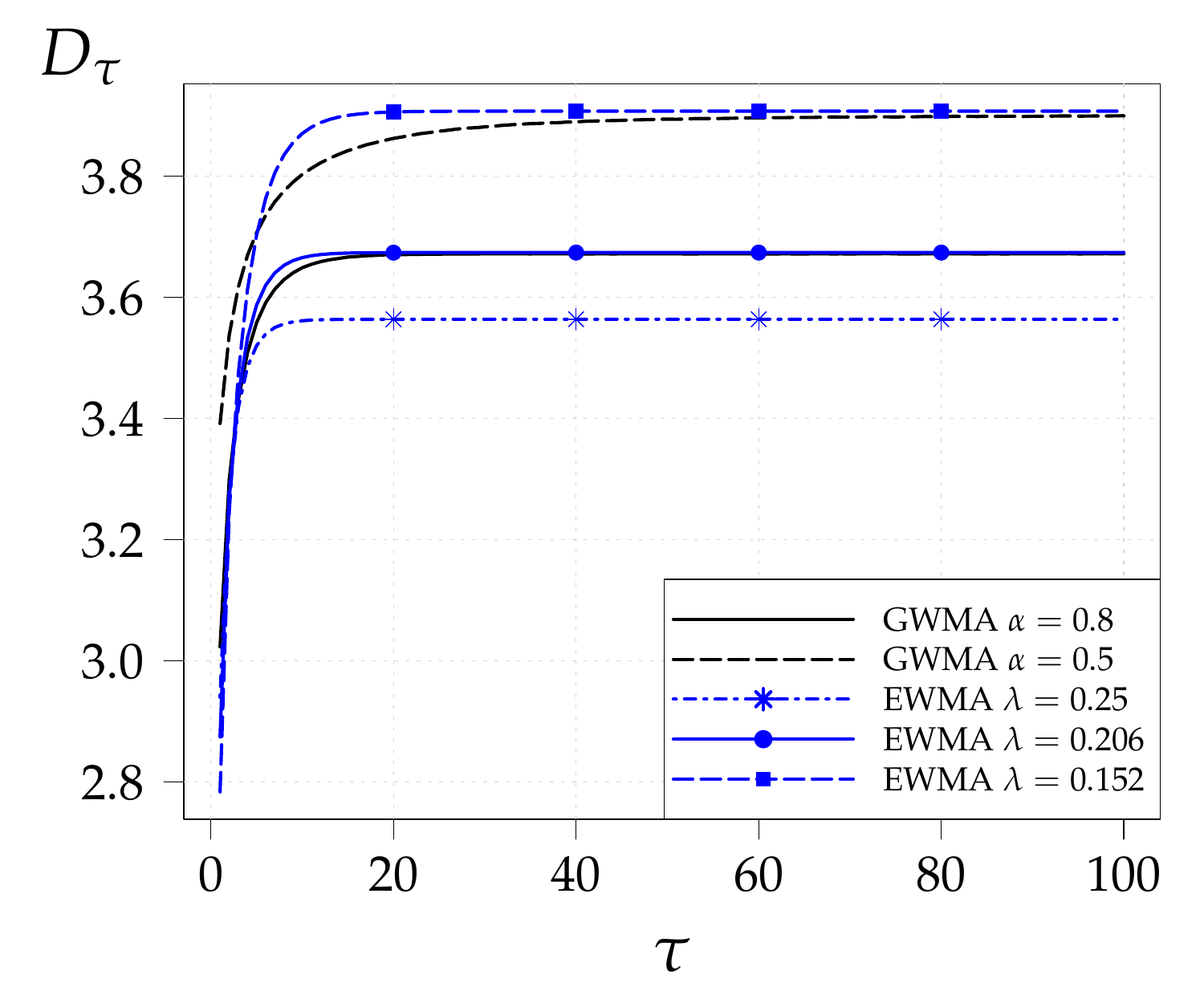} & 
  \includegraphics[width=.49\textwidth]{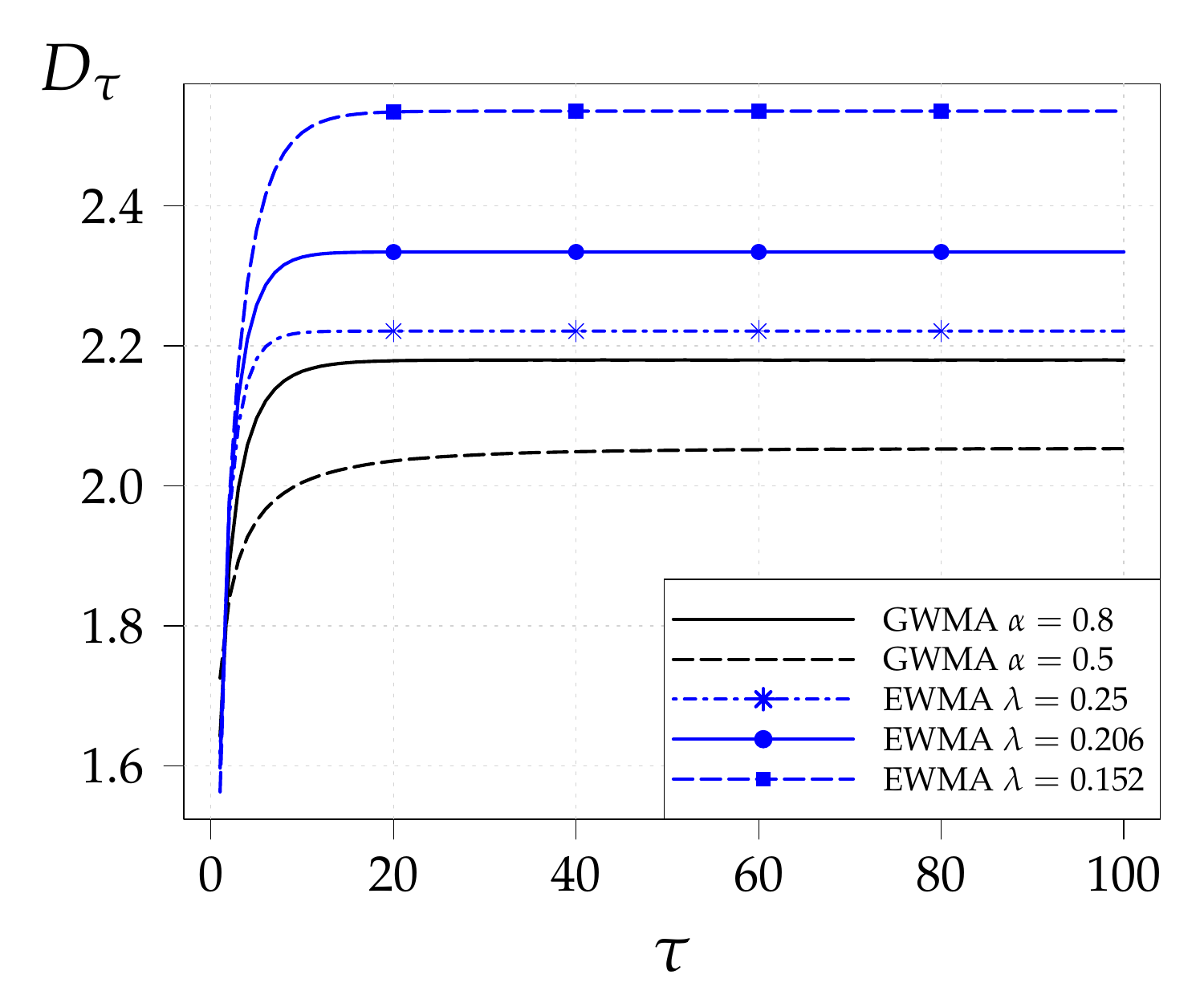}
\end{tabular}
\caption{Conditional expected delay $D_\tau$ for $q = 0.75$, $\alpha \in \{0.5,0.8\}$ and three EWMA designs,
medium and large change.}\label{fig:dtau_large}
\end{figure}

In the Figures~\ref{fig:dtau_small} and \ref{fig:dtau_large}, we display $\{D_\tau\}_{\tau=1}^{100}$
for the GWMA charts with $q = 0.75$, and $\alpha = 0.8$ and $ 0.5$.
In addition, we show three different EWMA chart profiles with
$\lambda = 1 - q = 0.25$, $ 0.206$ and $ 0.152$. Recall that
the latter two EWMA charts have the same asymptotic variance ($Q = Q_E$)
for $\alpha \in \{0.8, 0.5\}$. In Figure~\ref{fig:dtau_small},
we start with a small and a medium size change in the mean, 0.5 and 1 standard deviations.
For a shift of 0.5, the best performance is given for the EWMA chart with $\lambda = 0.152$,
whereas the two GWMA charts and the EWMA chart with $\lambda = 0.206$ perform similarly. The EWMA chart with $\lambda=0.25$ performs relatively poorly.
For the shift of 1 standard deviation, the two matched EWMA designs have the best performance, followed by the similarly behaving GWMA chart ($\alpha = 0.8$) and the third EWMA chart, and finally the worst performance is by the other GWMA chart.
We observe  that all $D_\tau$ values converge rather quickly and the two modified EWMA designs exhibit lower detection delays.

Turning to larger changes, magnitudes of 2 and 3 standard deviations, we recognize in Figure~\ref{fig:dtau_large} that the two corresponding GWMA/EWMA pairs ($\alpha=0.5$, $\lambda=0.152$ and $\alpha=0.8$, $\lambda=0.206$) behave nearly the same when the shift is of size 2 standard deviations, whereas the EWMA chart with ($\lambda=0.25$)
features now the smallest average delays. Only for the largest value of the shift (3 standard deviations) is the 
GWMA chart performance better than that of the three EWMA charts.
This somewhat surprising behavior is driven by the GWMA weighting patterns. For all past observations, the GWMA charts with $\alpha < 1$ weight
similarly to the EWMA chart with a decreased $\lambda$. The weight for the
most recent observation is larger, however, for the GWMA chart making it relatively more sensitive for large changes.

\begin{figure}[hbt]
\centering
\renewcommand{\tabcolsep}{0mm}
\begin{tabular}{cc}
  \scriptsize zero-state ARL &
  \scriptsize steady-state-state ARL \\[-1ex]
  \includegraphics[width=.49\textwidth]{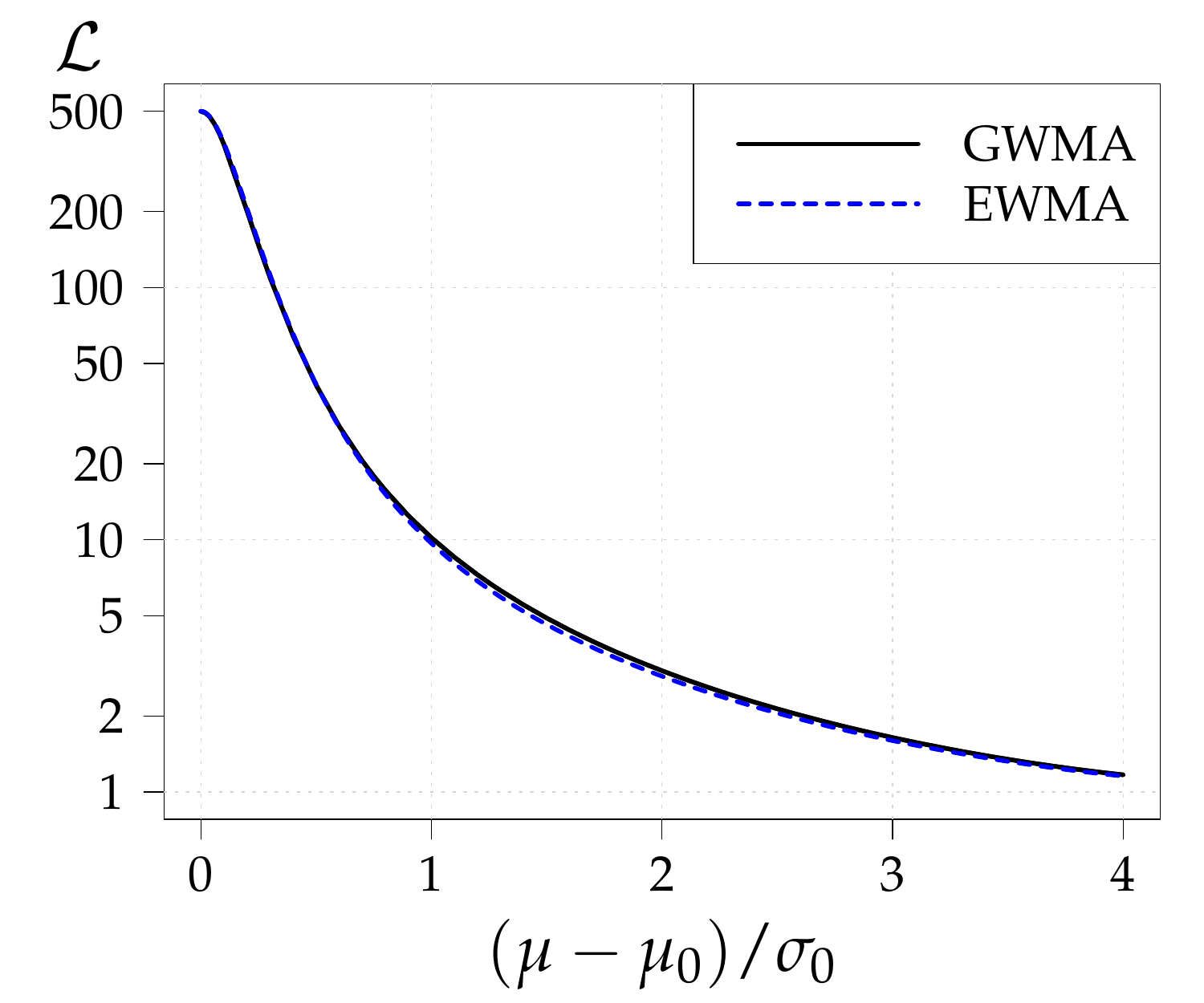} & 
  \includegraphics[width=.49\textwidth]{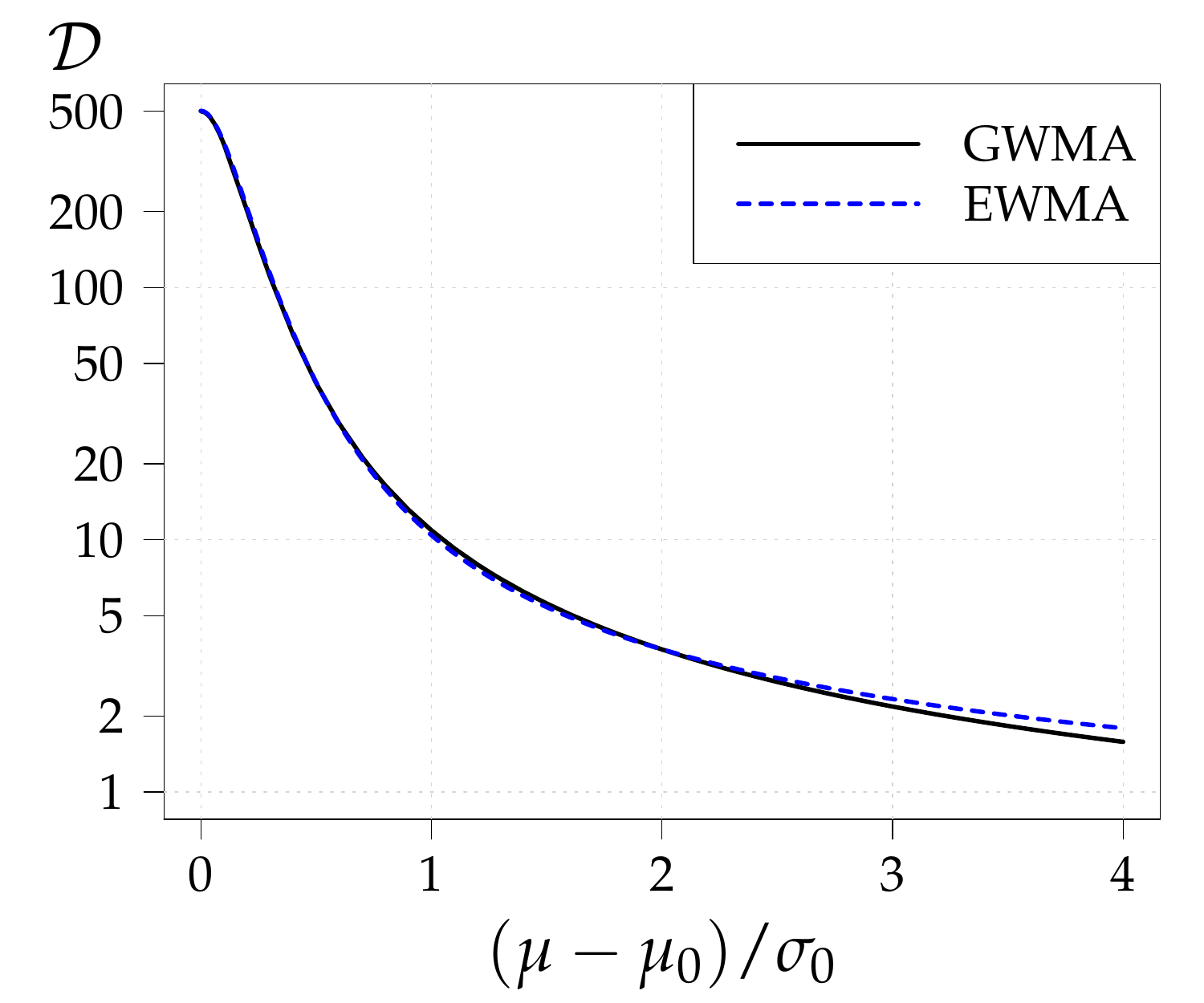}
\end{tabular}
\caption{Zero-state and steady-state ARL for
GWMA chart ($q = 0.75$, $\alpha=0.8$) and EWMA chart ($\lambda=0.206$).}\label{fig:ARL08}
\end{figure}

\begin{figure}[hbt]
\centering
\renewcommand{\tabcolsep}{0mm}
\begin{tabular}{cc}
  \scriptsize zero-state ARL &
  \scriptsize steady-state-state ARL \\[-1ex]
  \includegraphics[width=.49\textwidth]{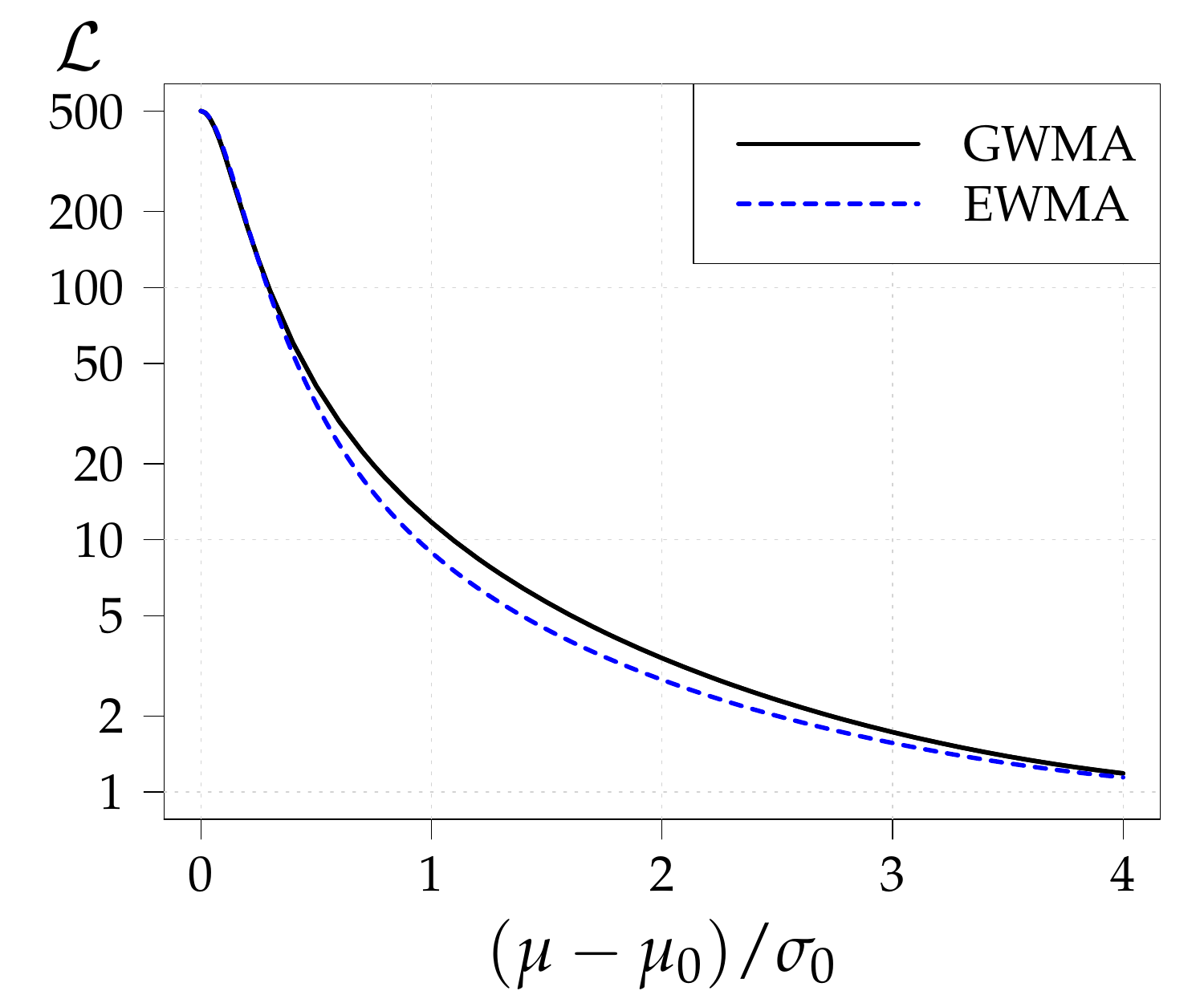} & 
  \includegraphics[width=.49\textwidth]{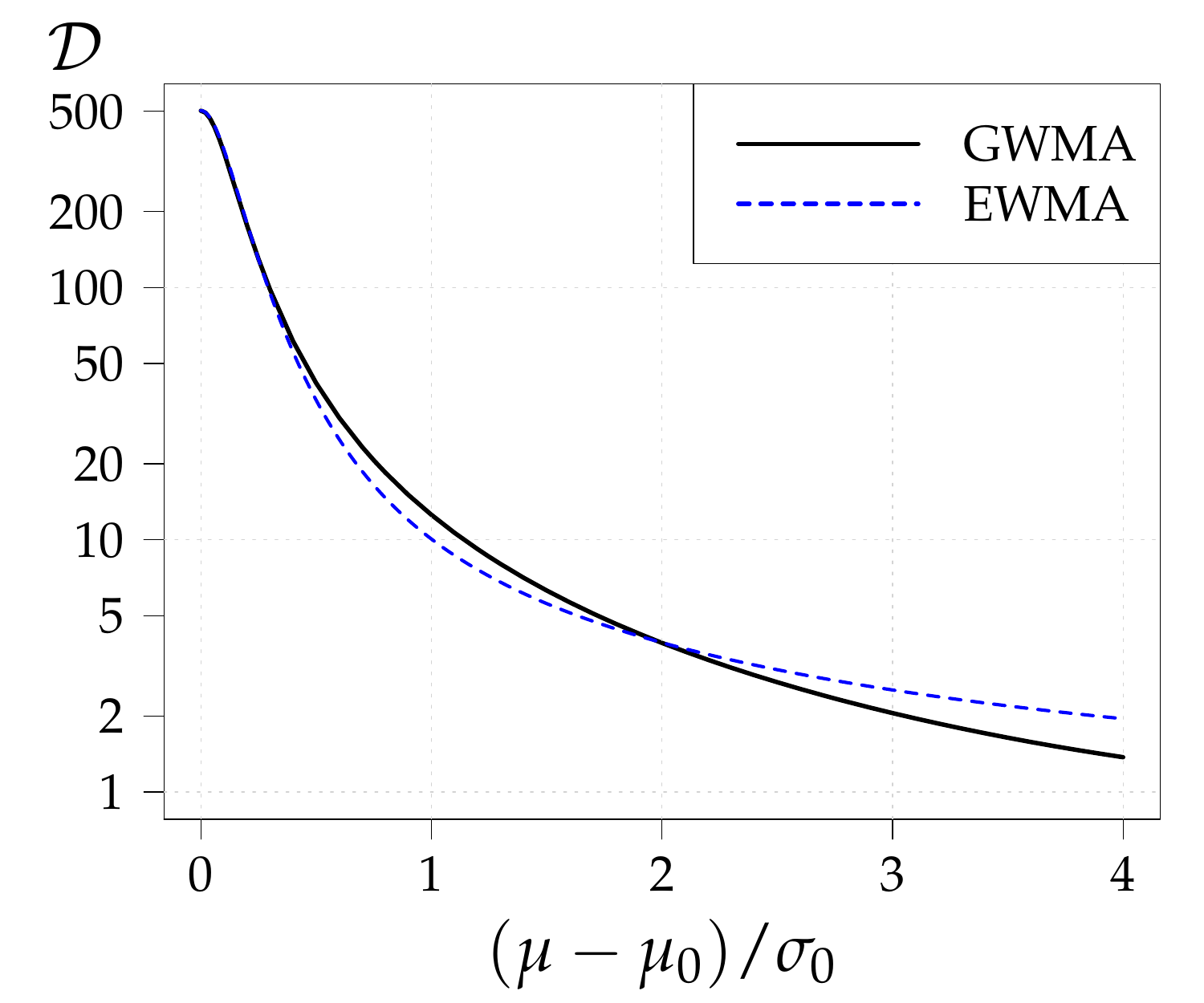}
\end{tabular}
\caption{Zero-state and steady-state ARL for
GWMA chart ($q = 0.75$, $\alpha=0.5$) and EWMA chart ($\lambda=0.152$).}\label{fig:ARL05}
\end{figure}

To provide an overview, we plot the zero-state ARL $\mathcal{L}$ and the conditional steady-state ARL $\mathcal{D}$ for the competing pairs  of charts against changes up to size 4 standard deviations. In Figure~\ref{fig:ARL08} we start
with ($\alpha=0.8$, $\lambda=0.206$).
In short, the GWMA chart with ($q = 0.75$, $\alpha=0.8$) and its
EWMA ($\lambda=0.206$) counterpart yield roughly the same performance curves.
Hence, the much less complicated and more established EWMA chart should be
preferred. Turning to the more pronounced GWMA weighting of past data, we
consider in Figure~\ref{fig:ARL05} the case
($q = 0.75$, $\alpha=0.5$).
Now, we discover slight advantages for the EWMA chart in terms
of the zero-state ARL for all shift magnitudes.
This remains valid for the steady-state ARL in case of
shifts sizes up to 2 standard deviations. Then the GWMA chart features slightly lower values.
Again, for practical applications one would prefer the EWMA chart.

\section{Conclusions}

We have pointed out that the GWMA control chart has serious and well-known computational disadvantages relative to the EWMA control chart. Its statistical performance is not better than that of an appropriately designed EWMA chart, so we see no justification whatsoever for its use. 

As a final note, it is implicitly assumed in our paper, and all papers on the GWMA approaches, that any process change,
however small, is to be detected quickly. If some process shifts are considered too small to be of concern, then we recommend
the approach of \cite{Wood:Falt:2019}.

\bibliographystyle{unsrtnat}
\bibliography{gwma}

\begin{thebibliography}{9}
\providecommand{\natexlab}[1]{#1}
\providecommand{\url}[1]{\texttt{#1}}
\expandafter\ifx\csname urlstyle\endcsname\relax
  \providecommand{\doi}[1]{doi: #1}\else
  \providecommand{\doi}{doi: \begingroup \urlstyle{rm}\Url}\fi

\bibitem[Sheu and Lin(2003)]{Sheu:Lin:2003}
Shey-Huei Sheu and Tse‐Chieh Lin.
\newblock The generally weighted moving average control chart for detecting
  small shifts in the process mean.
\newblock \emph{Quality Engineering}, 16\penalty0 (2):\penalty0 209--231, 2003.
\newblock \doi{10.1081/QEN-120024009}.

\bibitem[Sheu and Yang(2006)]{Sheu:Yang:2006a}
Shey-Huei Sheu and Ling Yang.
\newblock The generally weighted moving average control chart for monitoring
  the process median.
\newblock \emph{Quality Engineering}, 18:\penalty0 333--344, 2006.
\newblock \doi{10.1080/089821106007}.

\bibitem[Mabude et~al.(2021)Mabude, Malela-Majika, Castagliola, and
  Shongwe]{Mabu:EtAl:2021}
Kutele Mabude, Jean-Claude Malela-Majika, Philippe Castagliola, and Sandile~C.
  Shongwe.
\newblock Generally weighted moving average monitoring schemes: Overview and
  perspectives.
\newblock \emph{Quality and Reliability Engineering International}, 37\penalty0
  (2):\penalty0 409--432, 2021.
\newblock \doi{10.1002/qre.2765}.

\bibitem[Lucas and Saccucci(1990)]{Luca:Sacc:1990a}
James~M. Lucas and Michael~S. Saccucci.
\newblock Exponentially weighted moving average control schemes: Properties and
  enhancements.
\newblock \emph{Technometrics}, 32\penalty0 (1):\penalty0 1--12, 1990.
\newblock \doi{10.1080/00401706.1990.10484583}.

\bibitem[Knoth(2003)]{Knot:2003}
Sven Knoth.
\newblock {EWMA} schemes with non-homogeneous transition kernels.
\newblock \emph{Sequential Analysis}, 22\penalty0 (3):\penalty0 241--255, 2003.
\newblock \doi{10.1081/SQA-120025169}.

\bibitem[Knoth(2005)]{Knot:2005a}
Sven Knoth.
\newblock Fast initial response features for {EWMA} control charts.
\newblock \emph{Statistical Papers}, 46\penalty0 (1):\penalty0 47--64, 2005.
\newblock \doi{10.1007/BF02762034}.

\bibitem[Knoth(2021)]{Knot:2021}
Sven Knoth.
\newblock \emph{spc: Statistical Process Control -- Collection of Some Useful
  Functions}.
\newblock R Foundation for Statistical Computing, Vienna, Austria, 2021.
\newblock URL \url{https://cran.r-project.org/web/packages/spc/index.html}.
\newblock {R} package version 0.6.5.

\bibitem[Chakraborty et~al.(2017)Chakraborty, Human, and
  Balakrishnan]{Chak:Huma:Bala:2017}
Niladri Chakraborty, Schalk~W. Human, and Narayanaswamy Balakrishnan.
\newblock A generally weighted moving average chart for time between events.
\newblock \emph{Communications in Statistics -- Simulation and Computation},
  46\penalty0 (10):\penalty0 7790--7817, 2017.
\newblock \doi{10.1080/03610918.2016.1252397}.

\bibitem[Woodall and Faltin(2019)]{Wood:Falt:2019}
William~H. Woodall and Frederick~W. Faltin.
\newblock Rethinking control chart design and evaluation.
\newblock \emph{Quality Engineering}, 31\penalty0 (4):\penalty0 596--605, 2019.
\newblock \doi{10.1080/08982112.2019.1582779}.

\end{thebibliography}

\end{document}